\RequirePackage{lineno}
\documentclass[twocolumn,preprintnumbers,amsmath,amssymb,floatfix,showpacs]{revtex4-1}

\pdfoutput=1

\usepackage{graphicx}
\usepackage{atlasphysics}
\usepackage{subfigure}

\usepackage{hyperref}

\begin{document}

\preprint{CERN-PH-EP-2011-166}
\preprint{Submitted to Physical Review Letters}

\title{Measurement of the \ZZ production cross section and limits on anomalous neutral triple gauge couplings in proton-proton collisions at $\mathbf{\sqrt{s} =7}\TeV$ with the ATLAS detector}

\collaboration{The ATLAS Collaboration}

\date{November 25, 2011}

\begin{abstract}
A measurement of the \ZZ production cross section in proton-proton collisions
at $\sqrt{s} = 7\TeV$ using data recorded by the ATLAS experiment 
at the LHC is presented.  
In a data sample corresponding to an integrated luminosity of $1.02~\ifb$ collected in 2011,
$12$ events containing two \Z\ boson candidates decaying to electrons and/or muons are observed.
The expected background contribution is 
$0.3 \pm 0.3 \text{(stat.)} ^{+0.4}_{-0.3} \text{(syst.)}$ events. 
The cross section measured in a phase-space region with good detector acceptance and for dilepton masses within the range 66$\GeV$ to 116$\GeV$ is $\sigma_{ZZ\to \ll\ll}^\mathrm{fid} = 19.4^{+6.3}_{-5.2}\textrm{(stat.)}^{+0.9}_{-0.7}\textrm{(syst.)}\pm 0.7 \textrm{(lumi.)}~\fb$.
This result is then used to derive the total cross section for on-shell \ZZ production,
$\sigma_{ZZ}^\mathrm{tot}= 8.5^{+2.7}_{-2.3}\text{(stat.)}^{+0.4}_{-0.3} \text{(syst.)}{\pm0.3}\textrm{ (lumi.) }\pb$, which is consistent with the Standard Model
expectation of $6.5^{+0.3}_{-0.2}~\pb$ calculated at the next-to-leading order in QCD.
Limits on anomalous neutral triple gauge boson couplings are derived.
\end{abstract}


\maketitle

The production of pairs of \Z\ bosons at the LHC is of great interest since it provides an excellent 
opportunity to test the predictions of the electroweak sector of the Standard Model 
at the TeV energy scale; moreover it is the irreducible background to the 
search for the Higgs boson in the $\Hboson\ra ZZ$ decay channel.
In the Standard Model, \ZZ\ production proceeds at leading order (LO) via $t$-channel quark-antiquark  
interactions;
the $ZZZ$ and $ZZ\gamma$ neutral triple gauge boson couplings (nTGCs) are absent, 
hence there is no contribution from $s$-channel \qqbar\ annihilation at tree level. 
At the one-loop level, fermion triangles generate nTGCs of $\mathcal{O}(10^{-4})$~\cite{Gounaris:2000tb}. 
Many models of physics beyond the Standard Model
predict values of nTGCs at the level of $10^{-4}$ to $10^{-3}$~\cite{Ellison:1998}. 
The signature of non-zero nTGCs is an increase of the \ZZ cross section at high \ZZ invariant mass 
and high transverse momentum of the \Z\ bosons~\cite{Baur:2000ae}. 
\ZZ production has been studied in \epem\ collisions at 
LEP~\cite{Barate:1999jj,*Abdallah:2003dv,*Acciarri:1999ug,*Abbiendi:2003va,bib:LEPEW2006}
and in $p\overline{p}$ collisions at the Tevatron~\cite{bib:D0_ZZ1,bib:CDF_ZZ,*bib:D0_ZZ2,*bib:D0_ZZ3}.
No deviation of the measured cross section from the Standard Model expectation has been observed, 
and limits on anomalous nTGCs have been set~\cite{bib:LEPEW2006,bib:D0_ZZ1}. 

This letter presents the first measurement of \ZZ~\footnote{Throughout this paper \Z\ should be taken 
to mean $\Z/\gamma^{*}$.} production in proton-proton
collisions at a centre-of-mass energy $\sqrt{s}$ of 7 \TeV, and limits on the anomalous \mbox{nTGCs}. 
The cross section for on-shell \ZZ production (i.e.\ in the zero-width approximation) is
predicted at next-to-leading order (NLO) in QCD to be $6.5^{+0.3}_{-0.2}\ \pb$ \cite{Campbell:2011};
this includes a $\sim$6\% contribution from gluon fusion.
Candidate \ZZ events are reconstructed in
the $\ZZ\rightarrow\ll\ll$ decay channel, where $\ell$ can be an electron or muon. 
Although this channel constitutes only $\sim$0.5\% of the total \ZZ cross section, 
its final state with four high transverse-momentum, isolated leptons has a very high expected signal 
to background ratio of $\sim$30.

To reduce systematic uncertainties, the cross section is measured
within a phase-space that corresponds closely to the experimental acceptance;
this is termed the `fiducial' cross section.
The fiducial phase-space definition requires the invariant mass of both lepton pairs to be between 66 \GeV\ and 116 \GeV\ and all
four leptons to be within the pseudorapidity~\footnote{ATLAS uses a right-handed coordinate
system with its origin at the nominal interaction point in the centre of the detector and
the $z$-axis along the beam pipe. The $x$-axis points from the interaction point to
the centre of the LHC ring, and the $y$-axis points upwards. Cylindrical coordinates ($r$,$\phi$)
are used in the transverse plane, $\phi$ being the azimuthal angle around the beam pipe.
The pseudorapidity $\eta$ is defined in terms of the polar angle $\theta$ as
$\eta = - \ln\tan(\theta/2)$.} range $|\eta| < 2.5$ and have transverse 
momentum $\pT > 15$ \GeV. 
The four-momenta of all photons present after the simulation of the parton 
shower which are within 
$\Delta R \equiv \sqrt{\Delta \phi^2 + \Delta \eta^2} < 0.1$ 
of a lepton are summed into the four momentum of that lepton.
The total \ZZ cross section in the on-shell approximation
is obtained from the fiducial cross section 
using the known $\Z\rightarrow\ll$ branching ratio and a correction factor 
for the kinematic and geometrical acceptance.

Anomalous nTGCs for on-shell \ZZ production can be parameterized by two CP-violating ($f_4^V$) and two 
CP-conserving ($f_5^V$) complex parameters ($V = \Z, \gamma$) which are zero in the Standard 
Model~\cite{Baur:2000ae}.
To ensure partial-wave unitarity, a form-factor parameterization is introduced to cause
the couplings to vanish at high parton centre-of-mass energy  $\sqrt{\hat s}$: 
$f_i^V = f_{i0}^{V}/(1+\hat s/\Lambda^{2})^{n}$.
Here, $\Lambda$ is the energy scale at which physics beyond the Standard Model will be directly observable, $f^{V}_{i0}$ are the
low-energy approximations of the couplings, and $n$ is the form-factor
power. Following Ref.~\cite{Baur:2000ae}, $n = 3$
and $\Lambda = 2 \TeV$ are chosen, so that expected limits
are within the values provided by unitarity at LHC energies. The results with energy cutoff $\Lambda=\infty$ are also 
presented as a comparison in the unitarity violation scheme.

The ATLAS detector~\cite{bib:ATLASDetectorPaper} consists of inner tracking devices 
surrounded by a superconducting solenoid, electromagnetic and
hadronic calorimeters and a muon spectrometer with a toroidal magnetic field. The inner
detector, in combination with the 2\,T field from the solenoid, provides precision tracking of charged particles for $|\eta|<2.5$. It consists of a
silicon pixel detector, a silicon strip detector and a straw tube tracker that also provides 
transition radiation measurements for electron identification. The calorimeter system  covers the pseudorapidity range $|\eta| < 4.9$. 
It is composed of sampling calorimeters with either liquid argon (LAr) or scintillating tiles as the active media. 
In the region $|\eta|<2.5$ the electromagnetic LAr
calorimeter is finely segmented and plays an important role in electron identification.
The muon spectrometer has separate trigger and high-precision tracking chambers which provide muon
identification and measurement in $|\eta|<2.7$. 

A three-level trigger system selects events to be recorded for offline
analysis. The events used in this analysis were selected with  single-lepton triggers with nominal transverse
momentum thresholds of 20 \GeV\ for electrons and 18 \GeV\ for muons. 
The efficiencies of the single-lepton triggers have been determined as a function of lepton \pT\ using 
large samples of $\Z\ra\ll$ events. 
The trigger efficiency for events passing the offline selection described below is 99.9\% with an uncertainty 
of 0.1\%.

This measurement uses a data sample of proton-proton collisions at $\sqrt{s} = 7 \TeV$
recorded between February and June 2011. 
After data quality
requirements, the total integrated luminosity used in the analysis is \mbox{1.02 \ifb.} The 
integrated luminosity uncertainty is 3.7\%~\cite{bib:ATLASLumi2010,*ATLAS-lumi2011}.

Events are required to contain a primary vertex
formed from at least three associated tracks.
The vertex with the largest sum of the $\pT^2$ computed from the associated tracks is selected as 
the primary vertex.  

Signal events are characterized by four high-\pT, isolated electrons or muons, 
in three channels: \ee\ee, \mumu\mumu\ and \ee\mumu. Lepton candidates are required to be
consistent with originating from the primary vertex.
Muons are identified by matching tracks (or track segments) reconstructed in the muon spectrometer 
to tracks reconstructed in the inner detector~\cite{bib:ATLAS_W_Z}. 
Their momentum is calculated by combining the information from the two systems
and correcting for the energy deposited in the calorimeters. Only muons with $\pT >15\GeV$
and $|\eta| < 2.5$ are considered. 
In order to reject muons from the decay of heavy quarks, isolated muons are selected
by requiring
the scalar sum of the transverse momenta ($\Sigma \pT$) of other tracks with $\pt > 1 \GeV$ inside a cone of 
size $\Delta R = 0.2$
around the muon to be no more than 15\% of the muon \pT. 
The overall reconstruction, identification and isolation efficiency, measured in data using
a large sample of $\Z\rightarrow\mu^+\mu^-$ events, varies as a function of \pT\ from
92\% at 15 \GeV\ to 95\% at 45 \GeV.
  
Electrons are reconstructed from a cluster in the electromagnetic calorimeter matched to a track in the 
inner detector~\cite{bib:ATLAS_W_Z}. 
Electron candidates are required 
to pass the `medium' identification criteria described in Ref.~\cite{bib:ATLAS_W_Z}, 
to have a transverse momentum (measured in the calorimeter)
of at least 15 \GeV\ and a pseudorapidity of $|\eta| < 2.47$. They must be isolated, using the same 
criterion as for muons, calculating the $\Sigma\pT$ around the electron track.
Electron candidates within $\Delta R = 0.1$ of any selected muon
are rejected, and if two electron candidates are within $\Delta R = 0.1$ of each other the one with
the lower \pT\ is rejected.
The overall reconstruction, identification and isolation efficiency varies as a function of \pT\ from
63\% at 15 \GeV\ to 81\% at 45 \GeV.

Selected events are required to have exactly four leptons, and to have passed a 
single-muon or single-electron trigger. 
To ensure high trigger efficiency, at least one of these leptons must have $\pT > 20 \GeV\ (25 \GeV)$ for 
a muon (electron) and match to a muon (electron) reconstructed online by the trigger system within $\Delta R <$ 0.1 (0.15).

\begin{figure}[htbp]
 \begin{center}
  \includegraphics[width=0.5\textwidth]{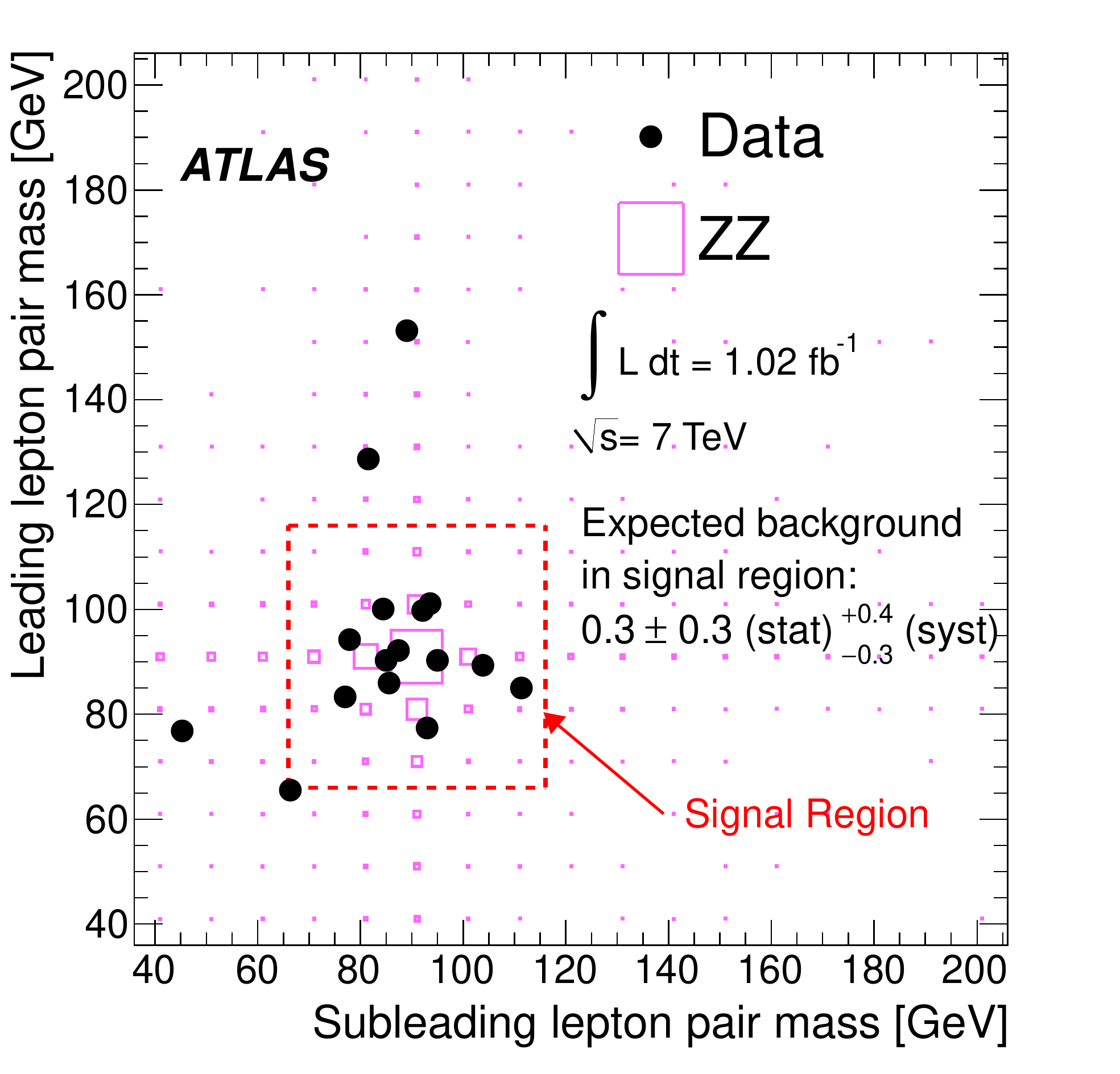}\hfill
  \caption{\small The mass of the leading lepton pair versus the mass of the subleading lepton pair.
  The events observed in the data are shown as solid circles and the \ZZ signal prediction from simulation as boxes.
  The large dashed box indicates the signal region defined by the requirements on the lepton-pair masses.
   }
 \label{fig:mass2d}
 \end{center}
 \end{figure}

Same-flavour, oppositely-charged lepton pairs are combined to form \Z\ candidates.
An event must contain two such pairs. 
In the \ee\ee\ and \mumu\mumu\ channels, ambiguities are resolved by choosing the pairing which
results in the smaller value of the sum of the two $|m_{\ll} - \mZ|$ values.
Figure~\ref{fig:mass2d} shows the correlation between the invariant mass 
of the leading (higher \pT) and the subleading (lower \pT) lepton pair.
The events cluster in the region where both masses are around \mZ.
Events are required to contain two \Z\ candidates with invariant masses satisfying 
$66 \GeV < m_{\ll} < 116 \GeV$.

The reconstruction efficiency for \ZZ events is determined from a detailed Monte Carlo simulation. 
The LO generator \pythia~\cite{pythia} with the MRST modified
LO parton density function (PDF) set~\cite{bib:MRST2007LOmod} 
is used to model  $pp \ra \ZZ \ra \ll\ll$ events,
where $\ell$ includes electrons, muons and $\tau$ leptons. 
The \pythia simulation includes the interference terms between the \Z\ and $\gamma^{*}$ diagrams; 
the mass threshold for the $\Z/\gamma$ boson is set to \mbox{12 \GeV}. 
The detector response is simulated~\cite{bib:ATLASMCPaper} with a program based on GEANT4~\cite{bib:geant4a}.
Additional inelastic $pp$ events are
included in the simulation, distributed so as to reproduce the number of collisions per 
bunch-crossing in the data. 
The simulation is also corrected with scale factors, and the lepton momentum resolution
adjusted, to reproduce the lepton reconstruction and identification
efficiencies measured in data. 

The overall efficiencies of the reconstruction and selection
criteria for events
generated within the fiducial phase-space are (40$\pm$3)\%, (79$\pm$2)\% and (57$\pm$2)\% for 
\ee\ee, \mumu\mumu\ and \ee\mumu\ respectively. 
The dominant systematic uncertainties arise from electron identification
(6.6\% in the \ee\ee\ final state, 3.1\% in the \ee\mumu\ final state) and 
from the muon reconstruction efficiency (2.0\% in \mumu\mumu\ and 1.0\% in \ee\mumu). 

Background to the \ZZ signal originates from events with a \Z\ (or \W) boson decaying to leptons
plus additional jets or photons ($W/Z+X$), from top-quark production and from other diboson
final states.
Such events may contain electrons or muons from the decay of heavy-flavoured hadrons, or muons from in-flight 
decay of pions and kaons; jets or photons may be misidentified as electrons. 
The majority of these background leptons are rejected by the isolation requirement. 

To estimate the background contribution from four-lepton events in which one lepton originates 
from a jet, a sample of events containing
three leptons passing all selection criteria plus one `lepton-like jet' is identified;
such events are denoted $\ell \ell \ell j$.
For muons, the lepton-like
jets are muon candidates that fail the isolation requirement. For electrons, the lepton-like
jets are clusters in the electromagnetic calorimeter matched to inner
detector tracks that fail either or both of the full electron selection and the isolation requirement. 
The events
are otherwise required to pass the full event selection, treating the lepton-like jet as if it
were a fully identified lepton. This event sample is dominated by $\Z + X$ events. 
The background is then estimated by scaling this control sample
by a measured factor $f$ 
which is the ratio of the probability for a jet to 
satisfy the full lepton criteria to the probability to satisfy the lepton-like jet criteria.
The background in which two selected leptons originate from jets is treated similarly, by 
identifying a data sample with two leptons and two lepton-like jets;
such events are denoted $\ell \ell j j$.
To avoid double counting in the background estimate,
and to take into account the expected \ZZ contribution in the control region, $N(\ZZ)$, 
the total number of background events $N(\rm{BG})$ is calculated as:
\begin{equation}
N({\rm BG}) = N(\ell\ell\ell j)\times f - N(\ell\ell jj)\times f^2 - N(\ZZ).
\end{equation}
The factor $f$ is measured in a sample of data selected with single-lepton triggers with criteria
applied to suppress isolated leptons from \W\ and \Z\ bosons,
and corrected for the remaining small contribution of true leptons using simulation. 
It is measured independently in \eta\ and \pT\ and the values combined assuming
they are uncorrelated.  
A similar analysis is performed on Monte Carlo simulation of background processes; the larger of the 
statistical uncertainty on $f$ determined from the data and the difference between data and simulation is taken 
as the systematic uncertainty in each \pT\ (or $\eta$) bin. 
This results in a systematic uncertainty which varies as a function of \pT\ 
from 57\% (85\%) at 15 \GeV\ to 55\% (77\%) at 45 \GeV\ for electrons (muons).

The numbers of expected and observed events after applying all selection criteria
are shown in Table~\ref{ta:selected_data_MC}. 
The expected number of signal events is determined from the \pythia simulation normalized to the NLO 
calculation using MCFM~\cite{Campbell:2011} with the MSTW2008~\cite{bib:MSTW2008} NLO PDF set. 
The normalization factor, calculated within the phase-space of the fiducial cross section measurement, 
is 1.41.
The expected numbers of signal events include contributions of 1.6\% from $\ZZ\ra\ll\ll$ events generated
outside the fiducial phase-space and 0.3\% from events where one of the \Z\ bosons decays to $\tau$
leptons.
Twelve \ZZ candidates are observed in data, with a background expectation of 
$0.3 \pm 0.3 \text{(stat.)} ^{+0.4}_{-0.3} \text{(syst.)}$,
corresponding to a p-value of $10^{-7}$ equivalent to a one-sided Gaussian significance of 5$\sigma$. 
In the four-muon channel 8 events are
observed where 
$3.3^{+0.4}_{-0.3}$
signal plus background events are expected. The probability of the expected number
fluctuating up to 8 or more is 3.2\%.  

\begin{figure*}[htbp]
 \begin{center}
 \subfigure[]{
 \includegraphics[width=0.43\textwidth]{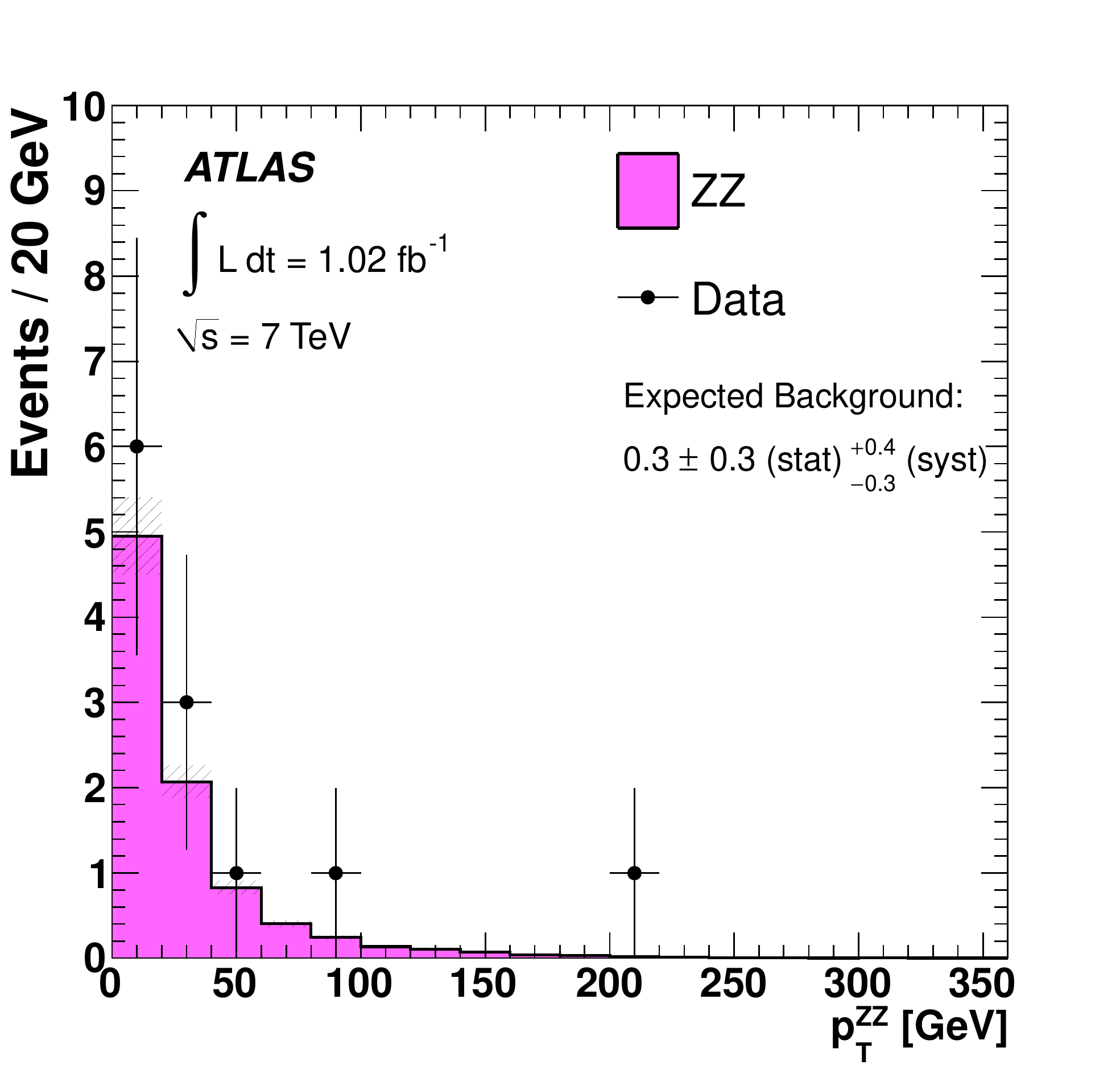}\hfill
 }
 \subfigure[]{
 \includegraphics[width=0.43\textwidth]{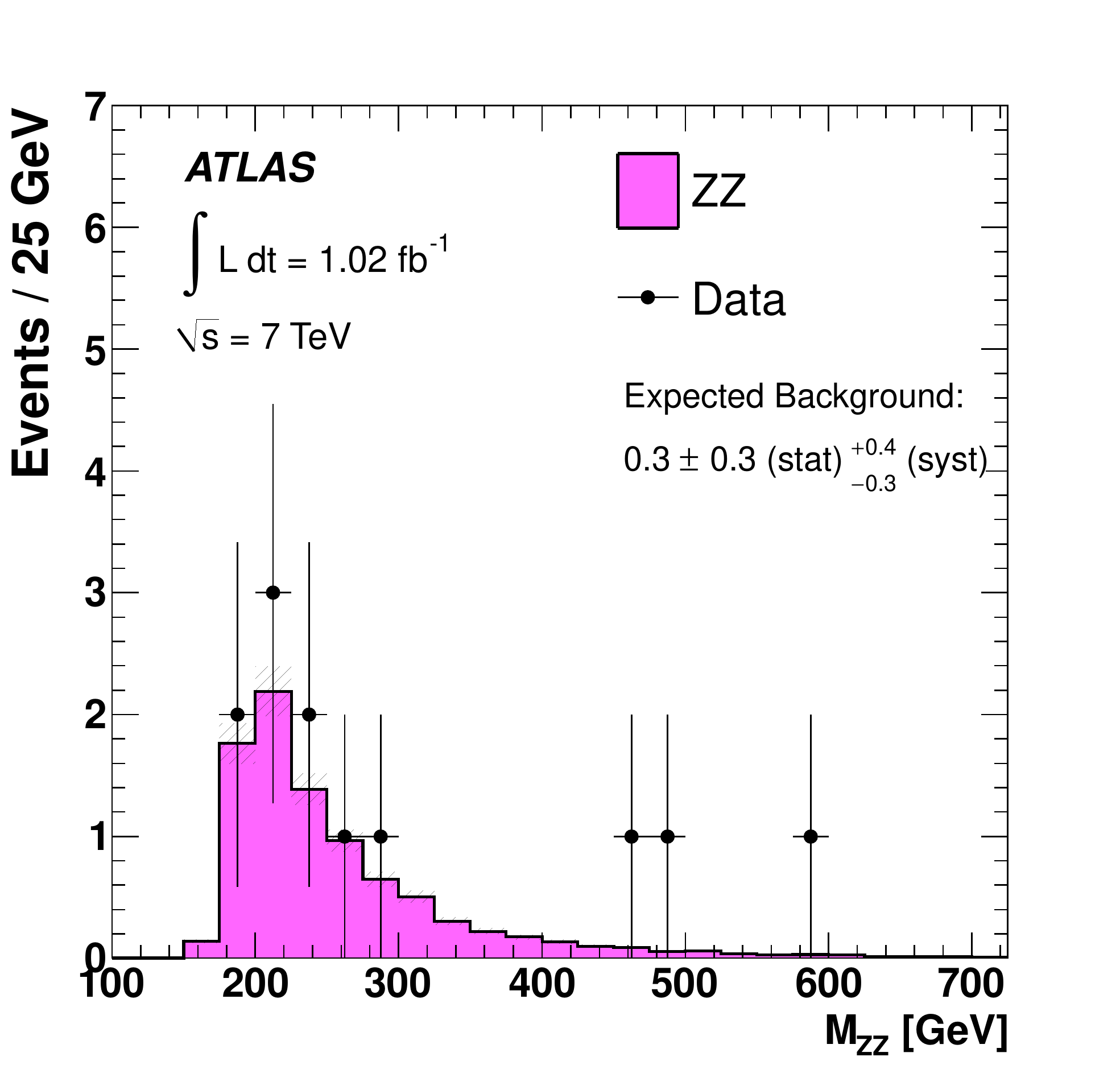}
 }
   \caption{\small (a) Transverse momentum $\pT^{\ZZ}$ and (b) invariant mass $M_{\ZZ}$ of the 
           four-lepton system for the selected events. The points represent the observed data and the 
           histograms show the signal prediction from simulation. The shaded band on each histogram
           shows the combined statistical and systematic uncertainty on the signal prediction. 
           The predicted number of background events from the data-driven background estimate is 
           indicated on the plot.
   \label{fig:kindists}
   }
 \end{center}
 \end{figure*}

\begin{table}
  \caption{\label{ta:selected_data_MC}
           Summary of observed events in the data, total background contributions and expected signal
	   in the individual four-lepton and combined channels. 
	   The quoted uncertainties represent 68.3\% confidence intervals; the first is statistical 
           while the second is systematic. The uncertainties on the
	   integrated luminosity (3.7\%) and the theoretical \ZZ cross section ($^{+4.7\%}_{-3.1\%}$) are not 
           included. 
          }
  \begin{ruledtabular}
  \begin{tabular}{lccc} 
  Channel     & Observed  &BG(data-driven)                    & Expected \ZZ           \\ \hline
  \ee\ee      & 2         &$0.01^{+0.03 +0.05}_{-0.01 - 0.01}$ & 1.53$\pm$0.03$\pm$0.10 \\
  \mumu\mumu  & 8         &$0.3 \pm 0.3 \pm0.3$                & 3.03$\pm$0.04$\pm$0.06 \\
  \ee\mumu    & 2         &$<0.01^{+0.03}_{-0.01}$             & 4.37$\pm$0.04$\pm$0.14    \\ \hline
  \ll\ll      &12         &$0.3 \pm 0.3 ^{+0.4}_{-0.3}$        & 8.9$\pm$0.1$\pm$0.3    \\ 
  \end{tabular}
  \end{ruledtabular}
\end{table}
The transverse momentum distribution and the invariant mass
distribution of the combined four-lepton system for the selected candidates
are shown in Fig.~\ref{fig:kindists}.

The \ZZ fiducial cross section is determined using a maximum likelihood fitting method to
combine the three four-lepton channels. The systematic uncertainties are included in the fitting
procedure as nuisance parameters.
The measured fiducial cross section
is:
\begin{eqnarray*}
\sigma_{ZZ\to \ll\ll}^\mathrm{fid} &=& 19.4^{+6.3}_{-5.2}\textrm{(stat.)}^{+0.9}_{-0.7}\textrm{(syst.)}\pm 0.7\textrm{(lumi.)}\mathrm{fb},
\end{eqnarray*}
where $\ll\ll$ refers to the sum of the \ee\ee, \ee\mumu\ and \mumu\mumu\ final states.
The total cross section is determined similarly, correcting for the known $\Z\to\ll$ branching
ratios and the acceptance of the fiducial phase-space. This acceptance, calculated at NLO 
using MCFM version 6.0 with the MSTW2008 PDF set, is $0.507\pm0.009$, 
where the error arises primarily from PDF uncertainties with a 1\% contribution from QED radiative corrections and off-shell $Z/\gamma^{*}$ effects evaluated from POWHEG BOX~\cite{Melia:2011tj}.
The measured value of the total on-shell \ZZ cross section is: 
\begin{eqnarray*}
\sigma_{ZZ}^\mathrm{tot} &=& 8.5^{+2.7}_{-2.3}\textrm{ (stat.) }^{+0.4}_{-0.3}\textrm{ (syst.) }{\pm 0.3}\textrm{ (lumi.) }\mathrm{pb}.
\end{eqnarray*}
The result is consistent within errors with the NLO Standard Model total cross section for this 
process of $6.5^{+0.3}_{-0.2}\ \pb$~\cite{Campbell:2011}.

\begin{table}
  \caption{\label{ta:TGCLimits}
           One dimensional 95\%\ confidence intervals for anomalous neutral gauge boson couplings, where 
           the limit for each coupling assumes the other couplings fixed at their Standard Model value.  
           Limits are presented for form factor scales of $\Lambda = 2$ \TeV\ and $\Lambda = \infty$ and include 
           both statistical and systematic uncertainties; the statistical uncertainties are dominant.
          }
  \begin{ruledtabular}
  \begin{tabular}{lcccc}  
     $\Lambda$  & $f_{40}^{\gamma}$ & $f_{40}^{Z}$ & $f_{50}^{\gamma}$ & $f_{50}^{Z}$\\
     \colrule
     $2$ \TeV  & $[-0.15, 0.15]$  & $[-0.12, 0.12]$ & $[-0.15,0.15]$ & $[-0.13, 0.13]$ \\  \hline  
     $\infty$  & $[-0.08, 0.08]$  & $[-0.07, 0.07]$ & $[-0.08,0.08]$ & $[-0.07, 0.07]$ \\  
  \end{tabular}
  \end{ruledtabular}
\end{table}

Limits on anomalous nTGCs are determined using the total number of observed events only. 
The \ZZ production yield dependency on couplings is parameterized using 
fully simulated events generated with SHERPA~\cite{Gleisberg:2008ta} subsequently reweighted using the leading-order matrix 
element~\cite{Baur:2000ae} within the framework of Ref.~\cite{Bella:2008wc}.
The reweighting procedure
uses simulated samples with Standard Model as well as non-Standard Model coupling values
to ensure adequate coverage of all kinematic regions. 
One dimensional 95\%\ confidence intervals for the anomalous nTGCs are determined using a maximum profile
likelihood fit to the observed number of events. The systematic errors are included as nuisance
parameters. 
The resulting limits for each coupling, determined assuming real couplings and with the other couplings 
fixed at their Standard Model value, are listed in Table~\ref{ta:TGCLimits}. 
The present results are dominated by statistical 
uncertainties: limits derived using statistical uncertainties alone differ from those in 
Table~\ref{ta:TGCLimits} by less than 0.01.
These limits are comparable with, or are more stringent than, those derived from measurements at 
LEP~\cite{bib:LEPEW2006} and the Tevatron~\cite{bib:D0_ZZ1}; it should be noted that
limits from LEP do not use a form factor, and those from the Tevatron use $\Lambda = 1.2\TeV$.

In summary, the \ZZ\ production cross section has been measured in proton-proton
collisions at $\sqrt{s}$ = 7 \TeV\ using the ATLAS detector. 
Both the fiducial cross section within the detector acceptance and the total cross
section have been determined. 
The latter is in agreement with the Standard Model expectation. 
Limits on anomalous nTGCs have been derived. 

We thank CERN for the very successful operation of the LHC, as well as the
support staff from our institutions without whom ATLAS could not be
operated efficiently.

We acknowledge the support of ANPCyT, Argentina; YerPhI, Armenia; ARC,
Australia; BMWF, Austria; ANAS, Azerbaijan; SSTC, Belarus; CNPq and FAPESP,
Brazil; NSERC, NRC and CFI, Canada; CERN; CONICYT, Chile; CAS, MOST and
NSFC, China; COLCIENCIAS, Colombia; MSMT CR, MPO CR and VSC CR, Czech
Republic; DNRF, DNSRC and Lundbeck Foundation, Denmark; ARTEMIS, European
Union; IN2P3-CNRS, CEA-DSM/IRFU, France; GNAS, Georgia; BMBF, DFG, HGF, MPG
and AvH Foundation, Germany; GSRT, Greece; ISF, MINERVA, GIF, DIP and
Benoziyo Center, Israel; INFN, Italy; MEXT and JSPS, Japan; CNRST, Morocco;
FOM and NWO, Netherlands; RCN, Norway; MNiSW, Poland; GRICES and FCT,
Portugal; MERYS (MECTS), Romania; MES of Russia and ROSATOM, Russian
Federation; JINR; MSTD, Serbia; MSSR, Slovakia; ARRS and MVZT, Slovenia;
DST/NRF, South Africa; MICINN, Spain; SRC and Wallenberg Foundation,
Sweden; SER, SNSF and Cantons of Bern and Geneva, Switzerland; NSC, Taiwan;
TAEK, Turkey; STFC, the Royal Society and Leverhulme Trust, United Kingdom;
DOE and NSF, United States of America.

The crucial computing support from all WLCG partners is acknowledged
gratefully, in particular from CERN and the ATLAS Tier-1 facilities at
TRIUMF (Canada), NDGF (Denmark, Norway, Sweden), CC-IN2P3 (France),
KIT/GridKA (Germany), INFN-CNAF (Italy), NL-T1 (Netherlands), PIC (Spain),
ASGC (Taiwan), RAL (UK) and BNL (USA) and in the Tier-2 facilities
worldwide.

\bibliography{ZZPaper}

\clearpage
\begin{flushleft}
{\Large The ATLAS Collaboration}

\bigskip

G.~Aad$^{\rm 48}$,
B.~Abbott$^{\rm 111}$,
J.~Abdallah$^{\rm 11}$,
A.A.~Abdelalim$^{\rm 49}$,
A.~Abdesselam$^{\rm 118}$,
O.~Abdinov$^{\rm 10}$,
B.~Abi$^{\rm 112}$,
M.~Abolins$^{\rm 88}$,
H.~Abramowicz$^{\rm 153}$,
H.~Abreu$^{\rm 115}$
% ...
\bigskip

% ...
$^{10}$ Institute of Physics, Azerbaijan Academy of Sciences, Baku, Azerbaijan\\
$^{11}$ Institut de F\'isica d'Altes Energies and Departament de F\'isica de la Universitat Aut\`onoma  de Barcelona and ICREA, Barcelona, Spain\\
% ...
$^{48}$ Fakult\"{a}t f\"{u}r Mathematik und Physik, Albert-Ludwigs-Universit\"{a}t, Freiburg i.Br., Germany\\
$^{49}$ Section de Physique, Universit\'e de Gen\`eve, Geneva, Switzerland\\
% ...
$^{88}$ Department of Physics and Astronomy, Michigan State University, East Lansing MI, United States of America\\
% ...
$^{111}$ Homer L. Dodge Department of Physics and Astronomy, University of Oklahoma, Norman OK, United States of America\\
$^{112}$ Department of Physics, Oklahoma State University, Stillwater OK, United States of America\\
% ...
$^{115}$ LAL, Univ. Paris-Sud and CNRS/IN2P3, Orsay, France\\
% ...
$^{118}$ Department of Physics, Oxford University, Oxford, United Kingdom\\
% ...
$^{153}$ Raymond and Beverly Sackler School of Physics and Astronomy, Tel Aviv University, Tel Aviv, Israel\\
% ...
\end{flushleft}

\end{document}